\documentclass[preprint,showpacs,preprintnumbers,amsmath,amssymb,superscriptaddress]{revtex4}

\usepackage{graphicx}
\usepackage{dcolumn}
\usepackage{bm}

\begin{document}

\preprint{APS/123-QED}

\title{Fermi surface topology and low-lying quasiparticle dynamics of parent Fe$_{1+x}$Te/Se Superconductor by orbital-polarization resolved ARPES}

\author{Y. Xia}
\affiliation{Joseph Henry Laboratories of Physics, Department of Physics, Princeton University, Princeton, NJ 08544}
\author{D. Qian}
\affiliation{Joseph Henry Laboratories of Physics, Department of Physics, Princeton University, Princeton, NJ 08544}
\affiliation{Department of Physics, Shanghai Jiao Tong University, Shanghai 200030, China}
\author{L. Wray}
\affiliation{Joseph Henry Laboratories of Physics, Department of Physics, Princeton University, Princeton, NJ 08544}
\affiliation{Lawrence Berkeley National Laboratory, University of California, Berkeley, CA 94305}
\author{D. Hsieh}
\affiliation{Joseph Henry Laboratories of Physics, Department of Physics, Princeton University, Princeton, NJ 08544}
\author{G.F. Chen}
\affiliation{Beijing National Laboratory for Condensed Matter
Physics, Institute of Physics, Chinese Academy of Sciences, Beijing, China}
\author{J.L. Luo}
\affiliation{Beijing National Laboratory for Condensed Matter
Physics, Institute of Physics, Chinese Academy of Sciences, Beijing, China}
\author{N.L. Wang}
\affiliation{Beijing National Laboratory for Condensed Matter
Physics, Institute of Physics, Chinese Academy of Sciences, Beijing, China}
\author{M.Z. Hasan}
\affiliation{Joseph Henry Laboratories of Physics, Department of Physics, Princeton University, Princeton, NJ 08544}
\affiliation{Princeton Center for Complex Materials, Princeton University, Princeton, NJ 08544}
\affiliation{Princeton Institute for the Science and Technology of Materials, Princeton University, Princeton, NJ 08544}

\date{\today}

\begin{abstract}

We report the first photoemission study of Fe$_{1+x}$Te - the host
compound of the newly discovered iron-chalcogenide
superconductors (maximum T$_c$ $\sim$ 27K). Our results reveal a pair of nearly electron-hole
compensated Fermi pockets, strong Fermi velocity renormalization
and an absence of a spin-density-wave gap. A shadow hole pocket is
observed at the "X"-point of the Brillouin zone which is
consistent with a long-range ordered magneto-structural
groundstate. No signature of Fermi surface nesting instability
associated with Q=($\pi$/2, $\pi$/2) is observed. Our results
collectively reveal that the Fe$_{1+x}$Te series is dramatically
different from the high T$_{c}$ pnictides and likely harbor unusual mechanism for superconductivity and magnetic order.


\end{abstract}

\maketitle


\maketitle

The discovery of superconductivity in the pnictides (FeAs-based
compounds) has generated interest in understanding the general
interplay of quantum magnetism, electronic structure and
superconductivity in iron-based layered compounds \cite{kami,
gfchen}. The recent observation of unusual superconductivity and
magnetic order in the structurally simpler compounds such as FeSe
and Fe$_{1+x}$Te are the highlights of current research
\cite{fese1, fese2, pressure, wangfete, sachdev}. The expectation is that
these compounds may provide a way to isolate the key ingredients
for superconductivity and the nature of the parent magnetically
order state which may potentially differentiate between vastly
different theoretical models \cite{ma, bernevig, subedi}. The
crystal structure of these superconductors comprises of a direct
stacking of tetrahedral FeTe layers along the c-axis bonded by weak van der Waals coupling.
Superconductivity with transition temperature up to 15K is
achieved in the Fe$_{1+x}$(Se,Te) series \cite{fese1, fese2} and
T$_c$ increases up to 27K under a modest application of pressure
\cite{pressure}. Density functional theories (DFT) predict that
the electronic structure is very similar to the iron-pnictides and
magnetic order in FeTe originates from very strong Fermi surface
(FS) nesting leading to the largest SDW gap in the series.
Consequently, the doped FeTe compounds are expected to exhibit
T$_c$ much higher than that observed in FeSe if
superconductivity would indeed be originating from the so called
($\pi$,0) spin fluctuations \cite{subedi}. These predictions
critically base their origin on the Fermi surface topology and the
band-structure details, however, no experimental results on the
Fermiology and band-structure exist on this sample class to this
date. Here we report the first angle-resolved photoemission spectroscopy
(ARPES) study of the Fe$_{1+x}$Te - the host compound of the
superconductor series. Our results reveal a pair of nearly
electron-hole compensated Fermi pockets, significant band
renormalization and a remarkable absence of the predicted large
SDW gap. Although the observed Fermi surface topology is broadly
consistent with the DFT calculations, no Fermi surface nesting
instability associated with the magnetic ordering wave vector was
observed. An additional hole pocket is observed in our data which
can be interpreted to be associated with a local-moment magneto-structural
groundstate in clear contrast to inter-band nesting. Our
measurements reported here collectively suggest that the FeTe
compound series is dramatically different from the parent compound
of the pnictide superconductors and may harbor novel forms of
magnetic and superconducting instabilities not present in the high
T$_{c}$ pnictides.

\begin{figure}
\includegraphics[width=0.8\textwidth]{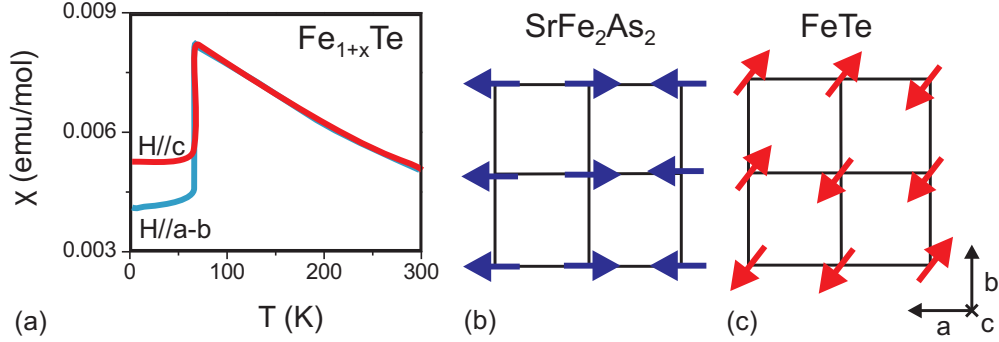}
\caption{\label{fig:spin} Magneto-structural transition and long-range order in Fe$_{1+x}$Te: (a)Temperature dependence of the magnetic
susceptibility \cite{wangfete}. (b) The spins in SrFe$_{2}$As$_{2}$ are collinear and \textbf{Q}$_{AF}$ points along the $(\pi, 0)$ direction. (c) The ordering vector, \textbf{Q}$_{AF}$ in FeTe, is rotated by 45$^{\circ}$ and points along the $(\frac{\pi}{2},\frac{\pi}{2})$ direction \cite{dai}. }
\end{figure}

Single crystals of Fe$_{1+x}$Te were grown using the Bridgeman technique. A mixture of
grounded Fe and Te powder was heated to 920$^{\circ}$C in an evacuated tube then
slowly cooled, forming single crystals. The iron concentration was measured by inductively-coupled plasma technique and x was determined to be less than 0.05. High-resolution ARPES
measurements were then performed using linearly-polarized 40eV
photons on beamline 10.0.1 of the Advanced Light Source at the
Lawrence Berkeley National Laboratory. The energy and momentum
resolution was 15meV and 2\% of the Brillouin Zone (BZ)
using a Scienta analyzer. The in-plane crystal orientation was
determined by Laue x-ray diffraction prior to inserting into the
ultra-high vacuum measurement chamber. The magnetic order below
65K was confirmed by DC susceptibility measurements (Fig-1). The
samples were cleaved \emph{in situ} at 10K under pressures of less
than $5\times 10^{-11}$ torr, resulting in shiny flat surfaces.
Cleavage properties were characterized by STM topography and by
examining the optical reflection properties.

\begin{figure}
\includegraphics[width=0.8\textwidth]{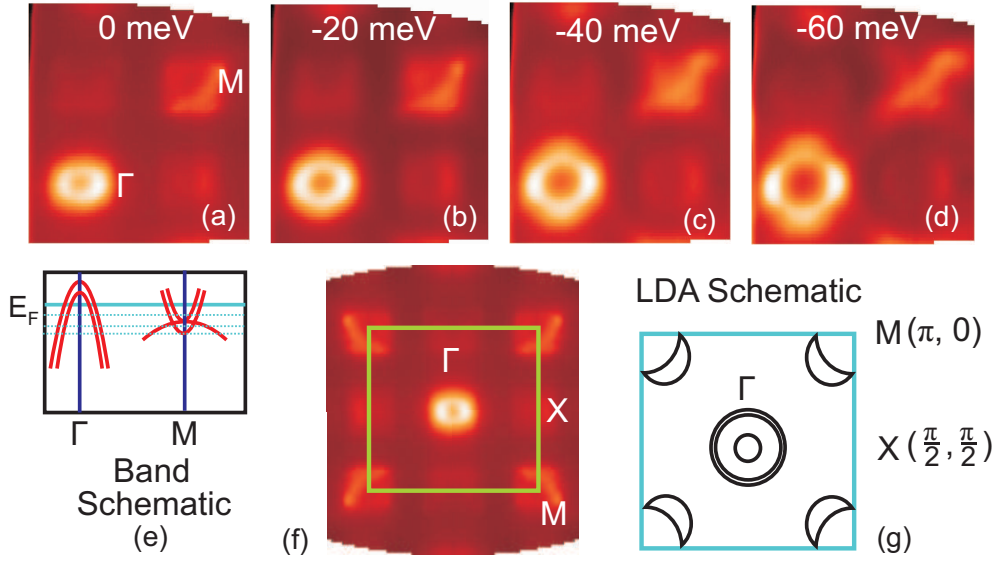}
\caption{\label{fig:fermi} \textbf{Fermi surface topology}: (a) Fermi surface topology of FeTe. (a-d) Connection to the underlying band-structure is revealed by considering the evolution of the density of states, n(k), as the chemical potential is rigidly lowered by 20, 40 and 60 meV. The Fermi surface consists of hole pockets centered at $\Gamma$, and electron pockets centered at M. An additional hole-like feature is observed at M in (b-d), which is attributed to a band lying below E$_F$ (see schematic in (e)). (e) presents a schematic of the band dispersions in the $\Gamma$-M direction forming the observed electron and hole pockets. A qualitative agreement is observed between the (f) the experimentally measured FS topology and (g) the LDA calculated \cite{subedi} FS although measured bands are strongly renormalized.}
\end{figure}

Figure~\ref{fig:fermi} presents the momentum dependence of the photoemission intensity n(k) at 10K at various binding
energies (0, -20, -40 and -60 meV), integrated over an energy
window of $\pm$ 5 meV at each binding energy. The non-zero spectral
intensity at the Fermi level ($\mu$) confirmed that the low
temperature state of Fe$_{1+x}$Te is metallic-like. At the Fermi
level, electrons are mostly distributed in one broad hole-like
pocket at $\Gamma$ and another similar size electron-like pocket
around M. To reveal the band shapes we present a gradual
binding-energy evolution of the band features and present the data
in a way to simulate the effect of rigidly lowering of the
chemical potential down to 60meV. The FS (at 0 meV) is seen to
consist of circular hole pockets centered at $\Gamma$, and
elliptical electron pockets at the M points. An elliptical-like
feature at the M point expands as the binding energy is increased,
suggesting that the associated band is hole-like. We will
subsequently show that this band lies below $\mu$ and the M-point
Fermi pockets are only electron-like (~\ref{fig:fermi}(e)). By
considering the dominant intensity patterns in panel-(a-d), apart
from a weak feature at "X"-point, the FS topology is similar to
that expected from the DFT calculations (~\ref{fig:fermi}(g))
\cite{subedi}. In addition, the area of the hole-like FS pocket at
$\Gamma$ and the electron-like FS pocket at M are approximately
equal in size suggesting nearly equal number of electron and hole
carrier densities in this material. Therefore, if the excess Fe
atoms are contributing to the carrier density it is likely to be
small and beyond our k-resolution of the experiment.

\begin{figure*}
\includegraphics[width=1.0\textwidth]{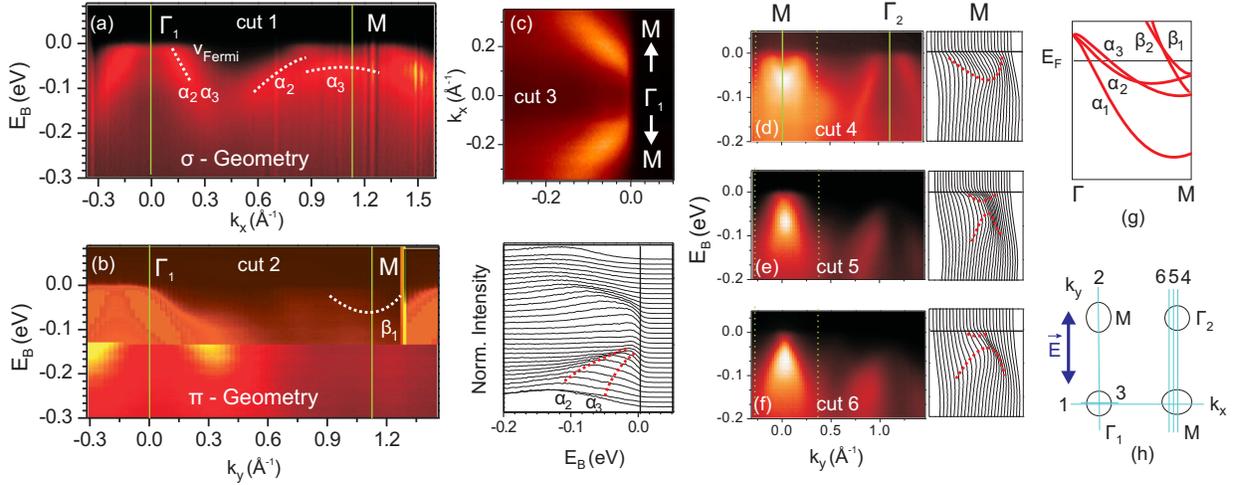}
\caption{\label{fig:cuts} \textbf{Low-lying band topology}: ARPES
spectra along the $\Gamma-M$ direction in the (a)$\sigma$ and
(b)$\pi$-scattering geometries. v$_F$ of the
quasiparticle band forming the $\Gamma$ hole pocket is calculated
from the dashed line. (c) The EDC of a high momentum resolution
scan through $\Gamma$ shows two hole-like bands crossing the
E$_F$. (d)-(f) Cuts along the M-$\Gamma$ direction through the
electron FS pocket show that the hole band near M remains at least
10meV below E$_F$. For each scan direction the corresponding EDC
is also presented, within the momentum range marked by green
dashed lines. Comparison with (g) a band dispersion schematic of
the LDA result \cite{subedi} shows a good agreement between
experiment and theory. (h) The directions of the six scans are
summarized and identified in the first BZ. The direction of the
electric field is shown.}
\end{figure*}

In order to systematically study the low lying energy band
structure, ARPES spectra are taken along different k-space cut
directions in the 2D Brillouin zone (BZ). Two different electron-photon
scattering geometries are used to ensure that all bands are imaged. In a
$\sigma$ scattering geometry, where the polarization vector is
along $k_y$ and the detector slit is along $k_x$, photoelectron signal is predominantly
from the $d_{xy}$ and $d_{yz}$ energy bands due to the
dipole emission matrix elements \cite{hsieh, qian}. Similarly, when the
detector slit is along $k_y$, the $\pi$-geometry, the
$d_{xz}$, $d_{z^2}$ and $d_{x^2-y^2}$ states are predominantly excited.
Figure ~\ref{fig:cuts} (a) and (b) present scans along the
$\Gamma-M$ direction in the $\sigma$ and $\pi$ geometries. In both
sets of spectra, one finds a broad hole-like band centered at
$\Gamma$. However, near M the scans taken at two different
geometries are drastically different. In cut 1, two hole-like
bands ($\alpha_2$, $\alpha_3$ as marked in panel (a)) are observed
to be approaching $\mu$. Under the $\pi$ geometry the band
emission pattern near M is dramatically different (see the
$\beta_1$ band in cut 2), while the $\alpha_2$ and $\alpha_3$ band
signals are significantly weaker. The polarization dependence
suggests that the $\beta_1$ band should have $d_{xz}$, $d_{z^2}$, or $d_{x^2-y^2}$ orbital character.

In order to fully resolve the broad band feature centered at
$\Gamma$, high resolution scans are performed along different
k-cut directions through the zone center. Figure
~\ref{fig:cuts}(c) presents one cut in the $\Gamma-M$ direction
inside the first zone, together with the corresponding energy
distribution curve (EDC). Two hole-like bands are resolved,
labeled $\alpha_2$ and $\alpha_3$. Since there are traces of multiple
bands near $M$, one might wonder whether the $\alpha_3$ band
crosses $\mu$ near the $M$ point, thus forming a hole-like  Fermi
pocket. To systematically investigate this, a series of spectra are
taken along the $\Gamma-M$ direction with $\pi$-geometry
(~\ref{fig:cuts}(d)-(f)) through multiple k-cuts intersecting the
$M$-point Fermi pocket. At $M$, cut-4 shows a strong electron-like
band forming the FS pocket. This band can be attributed to the
$\beta_1$ band (also observed in cut-2). As one moves away from M,
the band intensity becomes increasingly more hole-like, indicating
the emergence of the $\alpha_2$ band. Nevertheless, the hole band
lies completely and consistently below the Fermi level, with some
weak electron quasiparticle intensity above the band maximum. The
$\beta_1$ intensity becomes the weakest near the edge of the M
pocket (cut-6). But even at that location, the hole pocket lies at
least 10meV below the chemical potential. The result shows that
there are no hole pocket features in the FS near $M$ which
supports the interpretation that this material is nearly
electron-hole compensated. Additionally, one can map the observed
bands to the band structure estimated by DFT calculations
(schematic in ~\ref{fig:cuts}(g) \cite{subedi}). The calculation
finds three bands ($\alpha_1$-$\alpha_3$) crossing Fermi level
near $\Gamma$ and two ($\beta_1$, $\beta_2$) near M, forming the
electron and hole pockets. While the DFT calculated bands agree
fairly well with our data, within our resolution or the sample
surface roughness, we have not succeeded in fully resolving the
$\alpha_1$ band near $\Gamma$. The $\beta_2$ band, which forms the second M electron pocket, is observed in the Fermi surface topology (Fig.~\ref{fig:fermi}). The
overall band narrowing is about factor of 2 compared to the DFT
calculations highlighting the importance of correlation effects. This is also consistent with a small Fermi velocity (v$_{Fermi}$ $\sim$ 0.7 eV$\cdot$$\AA$) observed (Fig.3(a)).

\begin{figure}
\includegraphics[width=0.75\textwidth]{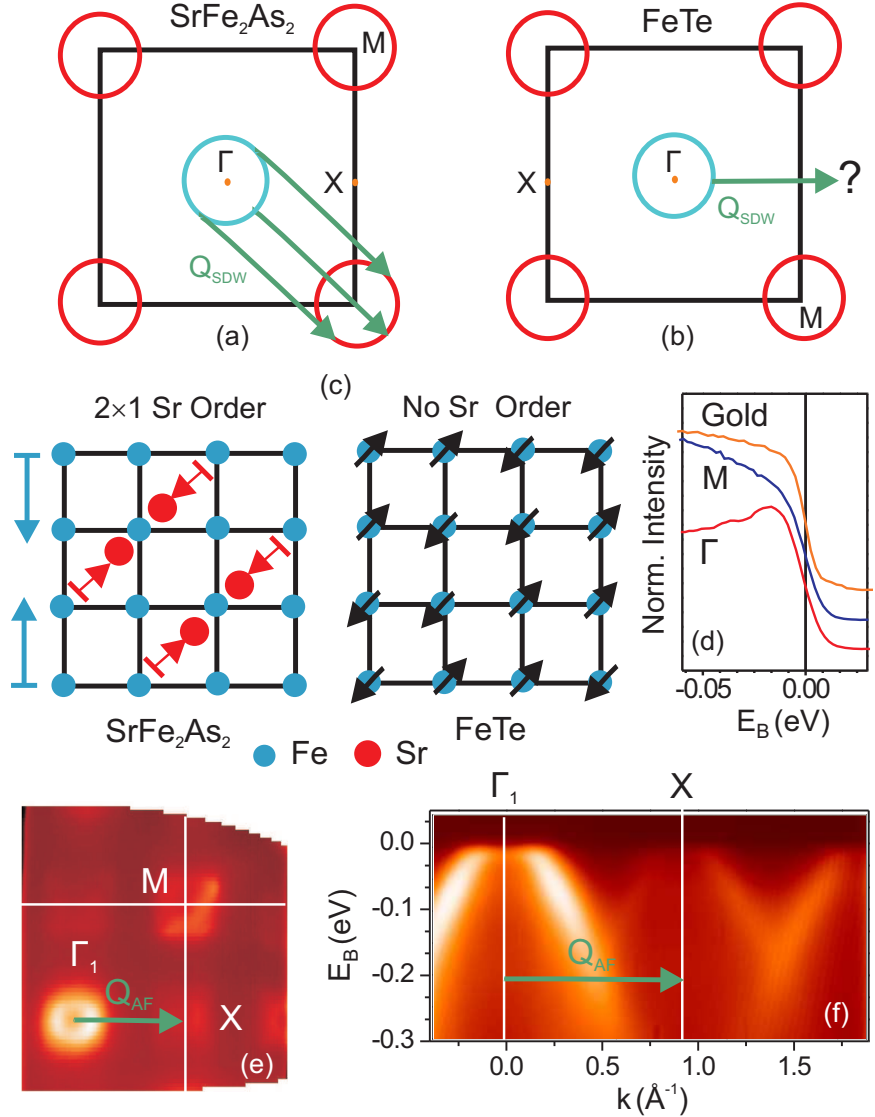}
\caption{\label{fig:fold} \textbf{Electronic structure and Magnetism}: While (a) the electron and hole pockets in SrFe$_2$As$_2$ can be kinematically nested by \textbf{Q}$_{AF}$=($\pi$,0), there exists (b) no FS pocket which
can be nested by \textbf{Q}$_{AF}$ = $(\frac{\pi}{2},\frac{\pi}{2})$ in the FeTe BZ. A FS pocket
observed at X in SrFe$_2$As$_2$ has been attributed to the (c)
$2\times 1$ surface order of the Sr atomic layer (red),
which occurs in addition to a weak bulk distortion (blue) across the
SDW transition. However, there are no additional Sr atoms between the
Fe layers in FeTe, so no Sr order is possible, although a weak bulk distortion is not excluded. (d) The EDCs measured at near the $\Gamma$ and M pockets exhibit no evidence of energy gaps. Nevertheless, (e) a weak FS pocket is observed at X, (f) corresponding to two hole-like bands dispersing towards the chemical potential which is consistent with a long-range local moment magneto-structural order. }
\end{figure}

We now revisit the details of the FS map (Fig.-2) and discuss a
weak feature observed at X=($\frac{\pi}{2},\frac{\pi}{2}$) between
two $\Gamma$ points (Fig. ~\ref{fig:fold}(e)). Such a feature is
not expected from the DFT calculations. A similar feature, whose
origin is debated, has been observed in AFe$_2$As$_2$
($\textbf{A}$=Ba, Sr), which is attributed to a $2 \times 1$ surface
reconstruction of the $\textbf{A}$ atom layer \cite{boyer, yin},
in addition to a weak bulk structural distortion. However, in FeTe
there are no additional atoms (such as Sr or Ba) between the Fe
layers and the crystal cleaves at a weak van der Waals bond
between two adjacent layers. Therefore, no strong 2$\times$1 long-range ordered
$\textbf{A}$-type reconstruction is expected except for a weak
bulk-like structural orthorhombicity tied to the magnetic order (magneto-structural effect). Recent neutron \cite{dai, bao} and
x-ray diffraction \cite{wu, takano} studies have shown that FeTe undergoes a bulk structural distortion from the tetragonal to weakly-monoclinic or orthorhombic phase near 65K, accompanied by long-range magnetic order
Q$_{AF}$=($\frac{\pi}{2},\frac{\pi}{2}$).

In the parent compound of the pnictide superconductors such as the
SrFe$_2$As$_2$ or BaFe$_2$As$_2$, the SDW vector
\textbf{Q}$_{SDW}=$($\pi$, 0) coincides with a Fermi surface
nesting vector connecting the hole-pocket at $\Gamma$ and the
electron pocket at M. Currently, it is believed that this
nesting is responsible for opening a gap in the low temperature
physical properties \cite{SFAgap, wangfete, hsieh}. In the case of FeTe,
while a nesting vector can indeed be drawn along
\textbf{Q}$_{SDW}=$($\pi$, 0) between a pair of electron and hole
pockets, all available neutron scattering measurements report that
the antiferromagnetic ordering vector is 45$^{\circ}$ away from
that in SrFe$_2$As$_2$, namely, in Fe$_{1+x}$Te,
\textbf{Q}$_{AF}=$($\frac{\pi}{2},\frac{\pi}{2}$). The ordering
shows a commensurate to incommensurate cross-over if the
concentration of excess iron, x, is increased. Another remarkable
difference is that the magnetic susceptibility is Curie-Weiss like
in FeTe suggesting that the magnetism is of local-moment origin. Within a local moment-like AFM long-range ordered state which also couples to a weak structural distortion one should expect relatively intense shadow Fermi surfaces along the Neel vector \textbf{Q$_{AF}$}. The X-point Fermi surface we observe thus can be related to the vector observed in neutron
scattering $X=\Gamma+$\textbf{Q}$_{AF}$. The weak shadow-like X
pocket FS might therefore arise from a band folding due to
long-range magnetic order. However, unlike SrFe$_2$Se$_2$, where
the $\Gamma$ electron pocket nests with the M hole pockets via
\textbf{Q}$_{SDW}=(\pi,0)$ \cite{zhao}, an analogous nesting
channel is unavailable at ($\frac{\pi}{2},\frac{\pi}{2}$) in the
FeTe (~\ref{fig:fold}(b)) clearly ruling out FS nesting as the
origin of magnetic order. In the absence of nesting FS gapping is
not expected in FeTe which is consistent with our results in Fig-2
and 3. A large low temperature specific heat value \cite{wangfete}
is thus consistent with our observation of a Fermi surface in the
correlated magneto-structurally ordered state. To examine the band dispersion character of the X pocket, figure ~\ref{fig:fold} (f) presents a spectra along the $\Gamma-X$ direction. Results show two bands dispersing towards
$\mu$, forming the hole pocket in the "X"-FS whose shapes are
indeed very similar to the bands that form the central $\Gamma$-pocket FS. Further evidence against nesting comes from the absence of a large gap at the overall low temperature electron distributions presented in Fig-2 and 3 (~\ref{fig:fold}(d)). Our results seem to suggest that the groundstate is a nearly electron-hole compensated semimetal. Absence of a gap is
consistent with recent bulk optical conductivity, specific heat
and Hall measurements in FeTe \cite{wangfete} (while most
measurements do report a gap in SrFe$_2$As$_2$ \cite{SFAgap}).

A recent density functional calculation \cite{singhfete} suggests
that the excess Fe is in a valence state near Fe$^+$ and therefore
donates electron to the system. Due to the interaction of the magnetic moment of excess Fe with the itinerant electrons of FeTe layer a complex magnetic ordering pattern is realized. In this scenario, excess Fe would lead to an enlargement of the electron pocket FS, however, within our resolution electron and hole Fermi surface pockets are measured to be of very similar
in size suggesting a lack of substantial electron doping due to
excess Fe. Finally, we note that a gapless yet long-range ordered
local-moment magneto-structural groundstate consistent with ARPES
and neutron data taken together is captured in both
first-principles electronic structure \cite{ma} and many-body spin
model \cite{bernevig} calculations. However, the broad agreement
of DFT calculations with experimental band-structure data except
for about a factor 2 renormalization is also remarkable. A
complete understanding of electron correlation, moment
localization and the true nature of the gapless antiferromagnetic
state would require further systematic ARPES study.

In conclusion, we have presented the first ARPES study of Fermi surface topology and band structure of Fe$_{1+x}$Te. Our results reveal a pair of \textit{nearly} electron-hole compensated Fermi pockets, significant renormalization and the remarkable absence of a spin-density-wave gap. The observed shadow hole pocket is consistent with a long-range ordered local moment-like magneto-structural groundstate whereas, most remarkably, no Fermi surface nesting instability associated with the antiferromagnetic order was observed. Contrary to band theory suggestions \cite{mj}, results collectively suggest that the Fe$_{1+x}$Te series is different from the undoped phases of the high T$_{c}$ pnictides and likely harbor a novel mechanism for superconductivity and quantum magnetism.

We thank P.W. Anderson, B. A. Bernevig, D. A. Huse, D.-H. Lee, Y. Ran, Z. Tesanovic, A. Vishwanath, and C. Xu for discussions. The synchrotron x-ray experiments at ALS/LBNL are supported by the U.S. DOE-BES (Contract No. DEFG02-05ER46200) and materials growth supported by
NSFC and CAS China (NLW).


\end{document}